\documentclass[runningheads]{llncs}
\usepackage[T1]{fontenc}
\usepackage{graphicx}
\usepackage{amsmath} 
\usepackage{amssymb}
\usepackage{enumitem}
\usepackage{booktabs}
\usepackage{multirow}
\usepackage{subcaption}
\usepackage{xcolor}

\title{MIL vs. Aggregation: Evaluating Patient-Level Survival Prediction Strategies Using Graph-Based Learning 
}
\author{Maria Rita Verdelho\inst{1}\orcidID{0000-0002-1407-8322} \and Alexandre Bernardino \inst{1}\orcidID{0000-0003-3991-1269} \and 
Catarina  Barata\inst{1}\orcidID{0000-0002-2852-7723}}
\authorrunning{MR Verdelho et al.}
\institute{Instituto Superior Técnico, 1049-001 Lisboa, Portugal }

\begin{document}
\maketitle              

\begin{abstract}

Oncologists often rely on a multitude of data, including whole-slide images (WSIs), to guide therapeutic decisions, aiming for the best patient outcome. However, predicting the prognosis of cancer patients can be a challenging task due to tumor heterogeneity and intra-patient variability, and the complexity of analyzing WSIs. These images are extremely large, containing billions of pixels, making direct processing computationally expensive and requiring specialized methods to extract relevant information. Additionally, multiple WSIs from the same patient may capture different tumor regions, some being more informative than others. This raises a fundamental question: Should we use all WSIs to characterize the patient, or should we identify the most representative slide for prognosis?

Our work seeks to answer this question by performing a comparison of various strategies for predicting survival at the WSI and patient level. The former treats each WSI as an independent sample, mimicking the strategy adopted in other works, while the latter comprises methods to either aggregate the predictions of the several WSIs or automatically identify the most relevant slide using multiple-instance learning (MIL). 

Additionally, we evaluate different Graph Neural Networks architectures under these strategies. We conduct our experiments using the MMIST-ccRCC dataset, which comprises patients with clear cell renal cell carcinoma (ccRCC). Our results show that MIL-based selection improves accuracy, suggesting that choosing the most representative slide benefits survival prediction.

\end{abstract}

\section{Introduction}
\label{sec1}

Whole-slide images (WSIs) are the gold standard for pathological diagnosis, prognosis, and treatment planning, offering a detailed view of tumor morphology \cite{Zhang}. Unlike other medical scans, WSIs are more difficult to collect, given their high storage cost. Therefore, public repositories such as the Cancer Genome Atlas (TCGA) and the Clinical Proteomic Tumor Analysis Consortium (CPTAC) \cite{tcga} have become essential resources for the development of computational approaches in WSI research. However, computational pathology faces critical challenges in the analysis of WSIs, primarily due to class imbalance, intra-patient variability, and the complexity of integrating multiple WSIs per patient.

WSIs are extremely large, high-resolution pathology images that capture intricate histological patterns at different spatial scales. Some slides contain diagnostically relevant tumor features, while others may be redundant or correspond to less informative tissue samples. For example, pathologists often include top and bottom slides to ensure that tumor margins are free of malignant cells, verifying that the tumor has been completely removed with a surrounding layer of healthy tissue. In TCGA, WSIs are labeled as follows: i) Diagnostic slides (DX): Typically the most relevant for tumor characterization; ii) Top slides (TS): Sections taken from the upper part of the tissue sample; and iii) Bottom slides (BS): Sections taken from the lower part of the tissue sample.
However, in CPTAC, where the number of slides per patient can be up to 9, such metadata is not available, making it unclear which slides best represent the tumor. This lack of labeling complicates the selection of data for specific tasks, such as survival prediction, as not all WSIs may be equally informative, and sets the tone for our research questions: (i) Should we use all available WSIs to characterize a patient, aggregating predictions through techniques like majority voting or one-dominance? (ii) Should we attempt to select a single, most representative WSI per patient, leveraging weakly supervised approaches like Multiple Instance Learning (MIL)?

\begin{table}[t]
\centering
\caption{Total number of images per category in TCGA and CPTAC used in the ccRCC benchmark.}\label{tab:total_images}
{\small
\begin{tabular}{@{}lcccccccccc|ccccccc@{}}
\cmidrule(l){2-18}
      & \multicolumn{10}{c}{\textbf{CPTAC} ($n=681$)} & \multicolumn{7}{c}{\textbf{TCGA} ($n=1790$)} \\ 
\cmidrule(l){2-11} \cmidrule(l){12-18}
\textbf{Label} & 21  & 22  & 23  & 24  & 25  & 26  & 27  & 28  & 29  & 30  & DX1  & DX2  & BS1  & BS2  & TS1  & TS2  & TSA   \\ 
\midrule
\textbf{Total} & 179  & 161  & 106  & 45  & 33  & 108  & 39  & 3  & 4  & 3  & 408  & 2  & 533 & 95  & 745  & 1  & 5   \\ 
\bottomrule
\end{tabular}}
\end{table}

We seek to answer these questions using the public MMIST-ccRCC dataset \cite{ccRCC} summarized in Table \ref{tab:total_images}. This set comprises 189 CPTAC and 429 TCGA patients with clear cell renal cell carcinoma (ccRCC), each associate to an average of 4 WSIs. However, it is important to mention that the problem that we are trying to handle is not exclusive to ccRCC and applies to other cancer datasets. We  compare two strategies for 12 month survival prediction:

\begin{enumerate}[label=\roman*)]
\item \textbf{WSI-Specific Prediction:} Each WSI is treated independently for survival prediction, allowing the model to assess tumor characteristics at the slide level without aggregating multiple WSIs per patient. This follows the common approach used in the literature.

\item \textbf{Patient-Specific Prediction:} We investigate whether it is more suitable to aggregate WSI predictions to derive a patient-Specific prediction or if is preferable to automatically select a single  scan followed by the patient prediction. For the former we will compare two strategies:
i) Majority Voting (MV) and ii) One-Dominance (1D). For the latter, we will adopt a MIL-based slide selection, where a model identifies the most representative WSI per patient to improve survival prediction, as described in  \cite{ccRCC}. 

\end{enumerate}

For each of the previous strategies we will conduct an extensive and systematic evaluation, supported by multiple runs of the same model. Additionally, we benchmark the performance of multiple graph neural network (GNN) architectures in the task of survival prediction. Graph-based representations have emerged as a powerful tool to model WSI data in a way that aligns with the biological structure of tissues, ensuring that prognostically relevant patterns, such as tumor microenvironment interactions and cellular organization, are effectively learned. However, a robust benchmark of the different GNN architectures is still missing for survival prediction, which this work seeks to mitigate.



\section{Related Work}

This section reviews approaches related to WSI-Specific Prediction, Patient-Specific Prediction, and GNNs for WSI analysis.  

\subsection{WSI Representation: MIL and Graph-Based Learning}

MIL is a widely used weakly supervised learning method for WSI analysis, where only slide-level labels are available. A WSI is divided into patches, grouped into bags, and labeled based on slide-level annotations \cite{Fatima}. The goal is to predict bag labels by aggregating information from instances \cite{Tarkhan}. MIL frameworks often use pooling functions for bag embeddings. Ilse \textit{et al.} \cite{Ilse} introduced an attention mechanism to assign importance scores to instances, computing a weighted average for bag representation. Li \textit{et al.} \cite{Li} proposed a dual-stream architecture, combining instance classification via max pooling and global instance aggregation. However, MIL struggles to capture spatial and contextual relationships within WSIs.

Graph-based methods address these limitations by incorporating spatial dependencies and contextual interactions. They have been applied to survival analysis \cite{Lii} \cite{Wang}, lymph node metastasis prediction \cite{Zhao}, and cancer staging \cite{Raju} \cite{Shi}. 
 Graph representations enable the modeling of intricate interactions within and across WSIs, offering a more holistic approach to understanding tumor heterogeneity. There are different strategies for Graph Representation:

\begin{enumerate}[label=\roman*)]
    \item \textbf{Cell-Graphs:} Introduced by \cite{Zhou}, these graphs model cellular interactions, where nodes represent cell nuclei, and edges capture relationships between them. However, cell-graphs often fail to account for the larger tissue architecture, limiting their utility for tasks requiring macroscopic context.
    \item \textbf{Tissue-Graphs:} The work in \cite{Anklin} proposed representing tissue regions as nodes and their interactions as edges. While these graphs capture broader spatial relationships, they overlook finer details of cellular interactions, leading to potential loss of critical information.
    \item \textbf{Patch-Graphs:} As a balance between fine-grained detail and computational efficiency, patch-graphs represent WSIs as graphs with patches as nodes. Chen \textit{ et al.} \cite{Chen} proposed constructing patch-graph convolutional networks, where edges are defined based on spatial proximity of patches, providing a scalable and effective way to preserve WSI spatial relationships. Unlike other works that explore feature similarity for edge definitions, their approach relied on spatial coordinates for adjacency, which simplifies graph construction. The work in \cite{Ritap} showed that Patch Graphs consistently outperformed other graph representations. 
\end{enumerate}

While graph-based methods have shown promise in tasks like cancer staging \cite{Raju} and survival analysis \cite{Liii}, most existing approaches focus on individual WSIs. Few have explored the integration of multiple WSIs into unified patient-level representations.

\subsection{Patient-Specific Prediction: Intra-Patient Variability in WSI Analysis}

Handling multiple WSIs for a single patient presents unique challenges. A growing body of work addresses this complexity:

\begin{enumerate}
    \item \textbf{Bag Aggregation in MIL:} Some MIL methods aggregate slide-level predictions to produce patient-level outputs. For example, \cite{Zhao} employed a meta-pooling strategy to combine predictions from multiple WSIs for lymph node metastasis prediction. However, these approaches often assume uniform relevance of slides, potentially diluting the importance of tumor-containing slides.
    \item \textbf{Instance Selection Across Slides:} Approaches such as \cite{Liii} select critical slides or patches using attention mechanisms to highlight tumor-relevant regions. These methods mitigate the influence of non-tumor slides but may fail to leverage inter-slide relationships fully.
    \item \textbf{Patient-Level MIL:} \cite{Wang} proposed a hierarchical MIL framework that aggregates features at the patch, slide, and patient levels. This hierarchical approach improves prediction accuracy by explicitly modeling relationships across slides but does not incorporate spatial relationships between WSIs.
\end{enumerate}

Building on these works, we compare One-Dominance (1D), Majority Voting (MV), and MIL-Based Selection to address intra-patient variability. Both 1D and Mil-Based Selection align with \textit{Instance Selection Across Slides}, selecting the most confident WSI and MV follows \textit{Bag Aggregation in MIL}, combining all WSIs for patient-level prediction.

\section{Methodology}

\begin{figure}[htbp]
    \centering
    \makebox[\textwidth]{\includegraphics[width=1.0\textwidth]{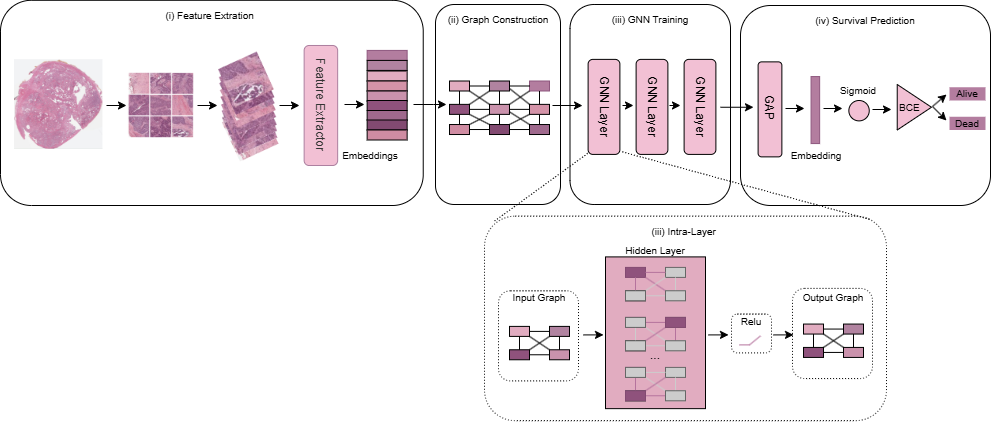}}
    \caption{Overview of the GNN survival prediction pipeline, including graph construction, multi-layer GNN processing, and survival prediction. The intra-layer block illustrates how each node updates its features based on its neighbors, with deeper layers leading to more homogeneous node representations.}
    \label{fig:gnn}
\end{figure}

This study proposes a graph-based framework for the analysis of multiple whole-slide images (WSIs) from patients suffering from clear cell renal cell carcinoma (ccRCC), aiming at predicting their survivability. The pipeline, illustrated in Figure \ref{fig:gnn}, consists of four main components: the Feature Extraction Module (\ref{subsec:FeatGraph}), the Graph Construction Module (\ref{subsec:Graph}), the GNN Training Module (\ref{subsec:Train}) and the Final Classification Module (\ref{subsec:class}). 

\subsection{Feature Extraction Module}
\label{subsec:FeatGraph}

Whole-slide images (WSIs) are gigapixel-scale pathology images that require efficient processing techniques due to their vast size. Directly applying deep learning models to entire WSIs is computationally infeasible, therefore it is needed to apply patch-based processing. In this paradigm, inspired by CLAM \cite{clam}, we present in Figure \ref{fig:gnn} in module (i) the Feature Extraction Module, where WSIs undergo segmentation to remove background artifacts. To improve segmentation accuracy, we fine-tune CLAM’s segmentation mask to align with CPTAC staining protocols, ensuring consistent patch-level feature extraction across datasets. This is followed by patching into non-overlapping $256 \times 256$ pixel regions at $20\times$ magnification, preserving high-resolution histological details while reducing computational costs.  Each patch is then processed with a ResNet-50 pretrained on ImageNet, generating a 1024-dimensional feature embedding. The resulting WSI representation is a feature matrix of size $X \in \mathbb{R}^{N_p \times 1024}$.

\subsection{Graph Construction Module}
\label{subsec:Graph}

Each WSI is represented as a patch graph, as depicted in Figure \ref{fig:gnn}, module (ii). Nodes correspond to patch-level features extracted during feature extraction, while edges encode spatial proximity relationships between patches. The adjacency matrix $A \in \mathbb{R}^{N_p \times N_p}$ (where $N_p$ is the number of patches) is defined as:

\begin{equation}
A_{i,j} =
\begin{cases}
1, & \text{if } j \in \mathcal{N}(i) \\
0, & \text{otherwise.}
\end{cases}
\end{equation}

where $\mathcal{N}(i)$ represents the set of neighboring nodes within a predefined spatial distance from node $i$. This ensures local histological structures are preserved while maintaining computational efficiency.

\subsection{GNN Training Module}
\label{subsec:Train}

The constructed graphs are processed using GNNs to capture spatial dependencies and histopathological relationships. Node features are iteratively aggregated from neighbors through multiple GNN layers, progressively refining representations. To prevent feature homogenization, we experimentally determine the optimal number of layers to be three, as deeper networks tend to mix features excessively, reducing node-level discrimination. Each GNN model updates node embeddings using a specific aggregation function. We evaluate three architectures:

\begin{enumerate}
    \item \textbf{Graph Convolutional Networks (GCN) \cite{GCN}:} Updates node embeddings via a spectral convolution operation:
    \begin{equation}
    H^{(l+1)} = \sigma \left( \tilde{D}^{-1/2} \tilde{A} \tilde{D}^{-1/2} H^{(l)} W^{(l)} \right),
    \end{equation}
    where $\tilde{A} = A + I$ is the adjacency matrix with self-loops; $\tilde{D}$ is the diagonal degree matrix, where $\tilde{D}_{ii} = \sum_j \tilde{A}_{ij}$; $H^{(l)}$ is the node feature matrix at layer $l$; $W^{(l)}$ is the trainable weight matrix; $\sigma(\cdot)$ is a non-linear activation function (ReLU).

    \item \textbf{Graph Attention Networks (GAT) \cite{GAT}:} Assigns learnable attention weights to neighboring nodes before aggregation:
    \begin{equation}
    \alpha_{ij} = \frac{\exp\left(\text{LeakyReLU}\left(a^{\top} \left[ W h_i \| W h_j \right]\right)\right)}{\sum_{k \in \mathcal{N}(i)} \exp\left(\text{LeakyReLU}\left(a^{\top} \left[ W h_i \| W h_k \right]\right)\right)},
    \end{equation}
    where $\alpha_{ij}$ is the learned attention coefficient for node $j$ relative to node $i$; $h_i, h_j$ are node embeddings; $W$ is the trainable weight matrix; $a$ is the attention vector

    \item \textbf{GraphSAGE (Full-Batch) \cite{GraphSage}:} Uses a sampling-based aggregation mechanism to balance efficiency and feature variability:
    \begin{equation}
    h_i^{(l+1)} = \sigma \left( W^{(l)} \cdot \text{AGGREGATE} \left( \{ h_i^{(l)}, h_j^{(l)} | j \in \mathcal{N}(i) \} \right) \right).
    \end{equation}
    where AGGREGATE is a neighborhood pooling function. Unlike GCN, GraphSAGE samples a subset of neighbors, preserving feature diversity.
    
    We employ Full-Batch (FB) GraphSAGE, which processes the entire graph at once, ensuring global feature aggregation, making it comparable to GCN and GAT.
\end{enumerate}

Each GNN consists of three layers, where node features are progressively updated through iterative neighborhood aggregation.

\subsection{Survival Prediction Module}
\label{subsec:class}

The classification module aggregates WSI predictions to determine patient-level labels (Figure \ref{fig:gnn}, module (iv)). A global average pooling (GAP) operation condenses graph embeddings, which are then processed by a fully connected layer with a sigmoid activation function:

\begin{equation} P(y) = \frac{1}{1 + \exp(-z)}, \end{equation}

where $z$ is the network's output. The model is optimized using binary cross-entropy (BCE) loss to predict if the patient is alive or dead.

\paragraph{Classification Scenarios:}
\begin{enumerate}
    \item \textbf{WSI-Specific Prediction:} 
    Each WSI is treated independently, with survival probabilities predicted per slide. Due to tumor heterogeneity, WSIs from the same patient may yield different outcomes.
    
    \item \textbf{Patient-Specific Prediction:} 
    Two strategies aggregate WSI predictions: (i) Majority Voting (MV), where the most frequent WSI prediction determines the final patient label, and (ii) One-Dominance (1D), selecting the WSI with the highest confidence score. Alternatively, a MIL-based slide selection approach \cite{ccRCC} identifies the most representative WSI per patient, improving survival prediction by prioritizing the most informative slide (Table \ref{tab:mil_images}).
\end{enumerate}

\begin{table}[t]
\centering
\caption{Total number of images per patient and per category in the ccRCC benchmark using the MIL-based slide selection.}\label{tab:mil_images}
{\small
\begin{tabular}{@{}lcccccccccc|ccccccc@{}}
\cmidrule(l){2-18}
      & \multicolumn{10}{c}{\textbf{CPTAC} ($n=189$)} & \multicolumn{7}{c}{\textbf{TCGA} (n=$429$)} \\ 
\cmidrule(l){2-11} \cmidrule(l){12-18}
\textbf{Label} & 21  & 22  & 23  & 24  & 25  & 26  & 27  & 28  & 29  & 30  & DX1  & DX2  & BS1  & BS2  & TS1  & TS2  & TSA   \\ 
\midrule
\textbf{Total} & 53  & 48  & 26  & 7  & 7  & 35  & 9  & 1  & 1  & 2  & 94  & 1  & 140 & 18  & 173  & 1  & 2   \\ 
\bottomrule
\end{tabular}}
\end{table}

These aggregation strategies aim to enhance prediction robustness by mitigating inconsistencies across multiple WSIs. As shown in Figure \ref{fig:gnn}, the slide-level predictions for a single patient may vary, and the final patient classification depends on the chosen aggregation method.

\section{Experimental Setup}
\subsection{Dataset}
The proposed framework is evaluated on the MMIST-ccRCC dataset \cite{ccRCC}, curated from the publicly available repositories TCGA, CPTAC, and TCIA. This dataset provides a comprehensive collection of WSIs for renal cell carcinoma analysis, focusing on the ccRCC. It comprises a total of 618 patients, with 497 assigned to the training set and 121 to the test set. Due to PyTorch variability, we employ 5 runs per model, ensuring robust and generalizable results across MMIST-ccRCC data splits. For evaluation metrics we used the balanced accuracy, recall and specificity.

\subsection{Implementation Details}
The proposed framework is implemented using PyTorch and PyTorch Geometric for graph-based computations. Models are trained using the Adam optimizer with a learning rate of 0.001 and weight decay of $5\times10^{-4}$.
The number of epochs is set to 
100, with early stopping based on validation loss with patience 10. A batch size of 32 is used for training.
To address the imbalance between tumor-positive and tumor-negative patients, class-weighted cross-entropy loss was used, with weights inversely proportional to class frequencies.
All experiments are conducted on a NVIDIA RTX3090.

\section{Results}
\label{results}
We evaluate two main strategies: (i) \textbf{WSI-Specific Prediction}, where each WSI is treated independently, and (ii) \textbf{Patient-Specific Prediction}, where WSI predictions are aggregated using One-Dominance (1D), Majority Voting (MV), or MIL-based slide selection to identify the most representative WSI per patient.

\subsection{WSI-Specific Prediction: Treating WSIs Independently}
\begin{table*}[t]
    \centering
    \renewcommand{\arraystretch}{1.2} 
    \setlength{\tabcolsep}{5pt} 
    \caption{Performance comparison of different GNN architectures using WSIs independently for survival prediction on MMIST-ccRCC.}
    \begin{tabular}{|c|c|c|c|}
        \hline
        \textbf{Model} & \textbf{Bacc} & \textbf{Recall} & \textbf{Specificity} \\
        \hline
        GCN & $0.6873 \pm 0.0061$ & $0.8457 \pm 0.0157$ & $0.5288 \pm 0.0254$ \\
        GAT 1 head & $0.6917 \pm 0.0118$ & $0.8409 \pm 0.0490$ & $0.5425 \pm 0.0718$ \\
        GAT 2 heads & $0.6866 \pm 0.0181$ & $0.7951 \pm 0.0714$ & $0.5781 \pm 0.0609$ \\
        GraphSAGE FB(mean) & $\textbf{0.7003} \pm 0.0089$ & $\textbf{0.8608} \pm 0.0560$ & $0.5397 \pm 0.0479$ \\
        GraphSAGE FB(max) & $0.6998 \pm 0.0135$ & $0.7995 \pm 0.0572$ & $\textbf{0.6000} \pm 0.0401$ \\
        \hline
    \end{tabular}
    \label{tab:wsi_prediction}
\end{table*}

We first evaluate the performance of different GNN architectures by treating each whole-slide image (WSI) independently. This WSI-specific approach allows us to analyze how well each model performs at the slide level before incorporating patient-level aggregation strategies.

Table \ref{tab:wsi_prediction} summarize the results of various GNN architectures under the WSI-specific strategy. GraphSAGE FB (mean pooling) demonstrates the highest balanced accuracy ($0.7003$), highlighting its strong generalization capability. GCN struggles with specificity ($0.5288$), indicating a tendency toward higher false-positive rates, which could be problematic in clinical settings where reducing false alarms is crucial. This could be attributed to its neighborhood averaging mechanism, which might not be optimal for discriminating between tumor and non-tumor regions. GAT shows a notable improvement in specificity with multiple attention heads. While GAT with a single head achieves $0.5425$ specificity, using two heads improves specificity to $0.5781$, suggesting that multi-head attention enhances feature aggregation. Finally, GraphSAGE FB (max pooling) performs slightly lower than its mean pooling counterpart, achieving a balanced accuracy of $0.6998$. This indicates that while max pooling can help emphasize the most salient features, mean pooling provides a more stable aggregation approach.



\subsection{Patient-Specific Prediction}

\begin{figure}[h]
    \centering

    \begin{subfigure}[b]{0.48\textwidth}
        \centering
        \includegraphics[width=\textwidth]{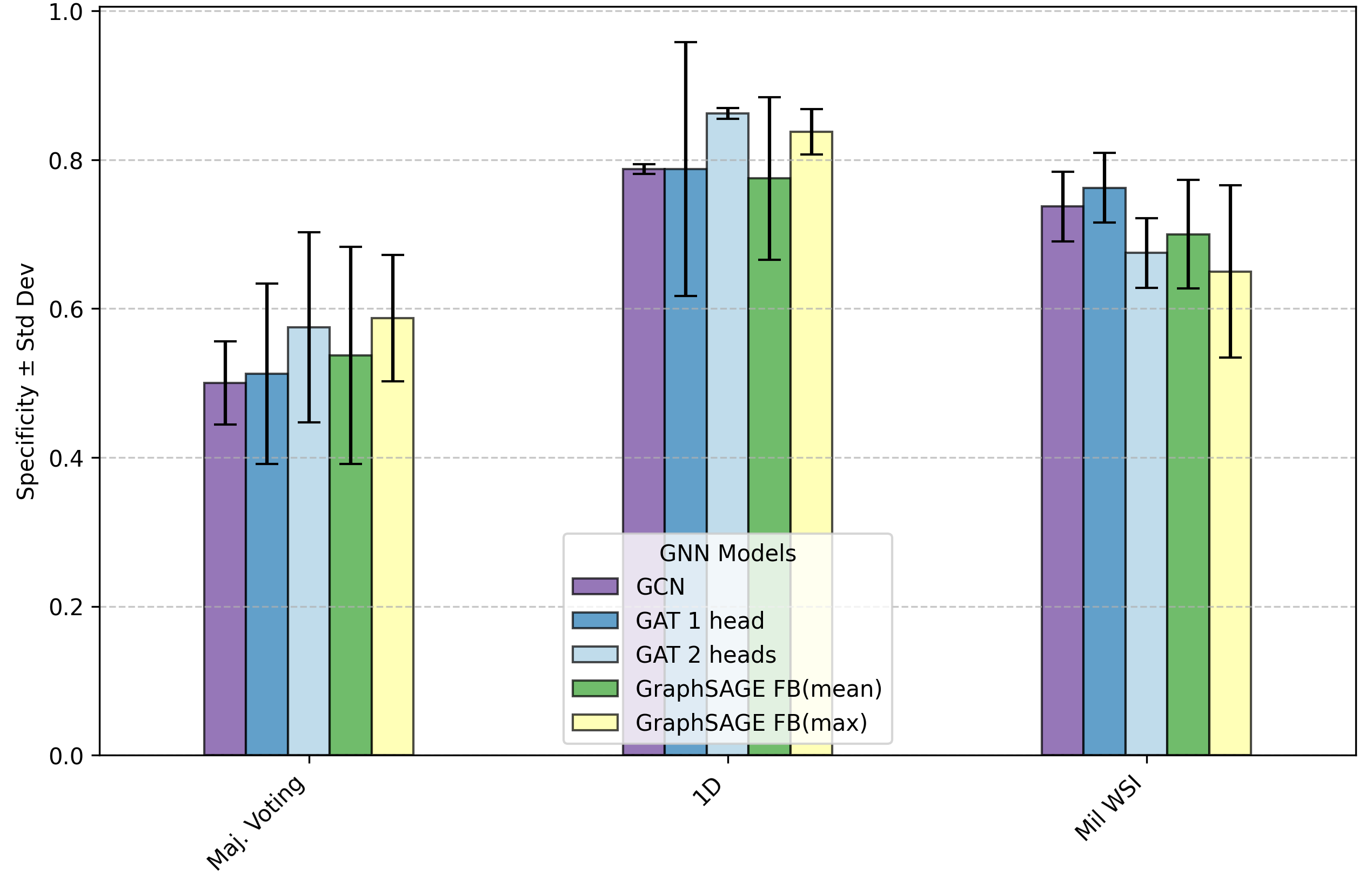}
        \caption{Specificity}
        \label{fig:bacc}
    \end{subfigure}
    \hfill
    \begin{subfigure}[b]{0.48\textwidth}
        \centering
        \includegraphics[width=\textwidth]{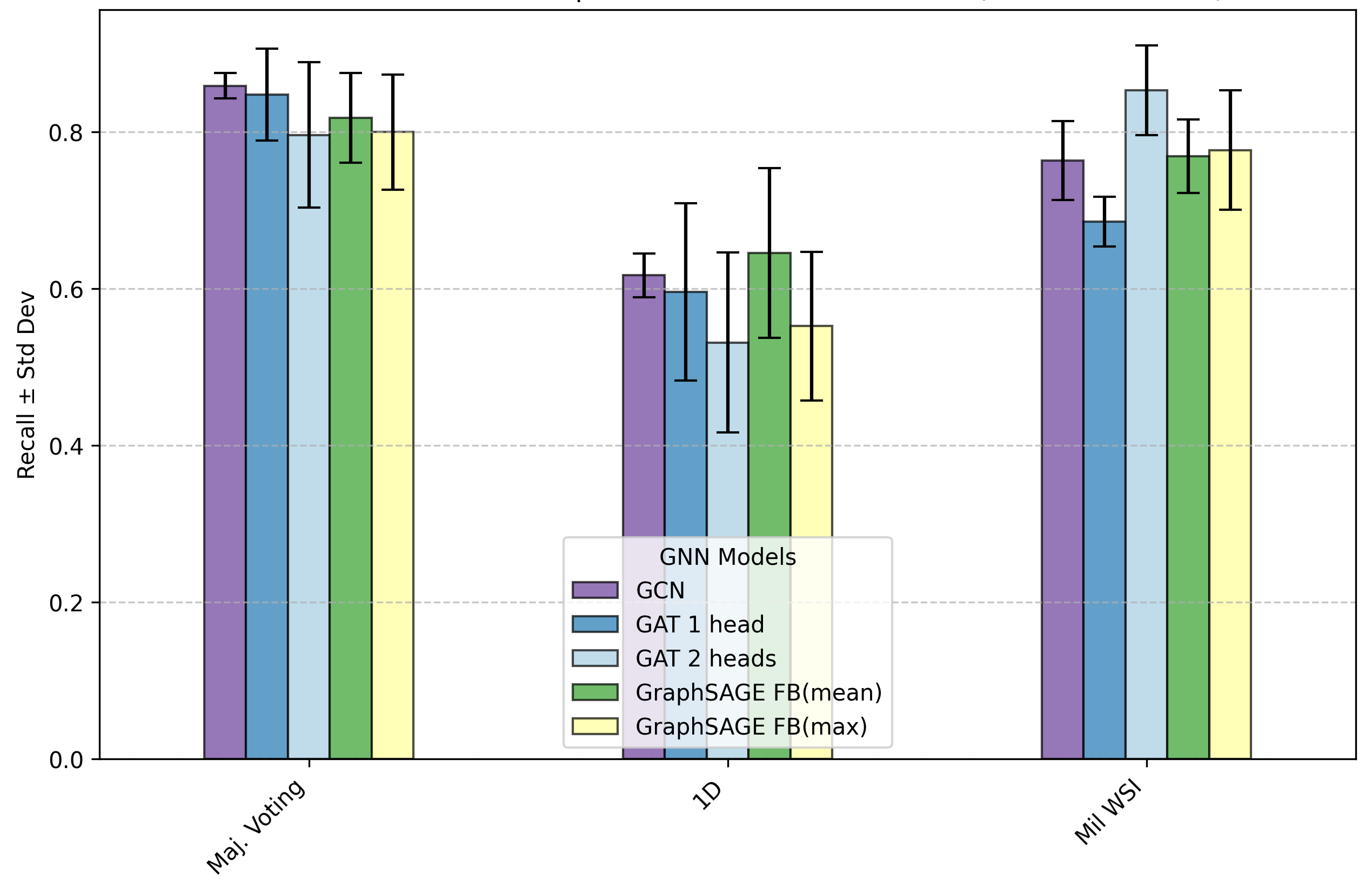}
        \caption{Recall}
        \label{fig:recall}
    \end{subfigure}

    \begin{subfigure}[b]{0.49\textwidth}
        \centering
        \includegraphics[width=\textwidth]{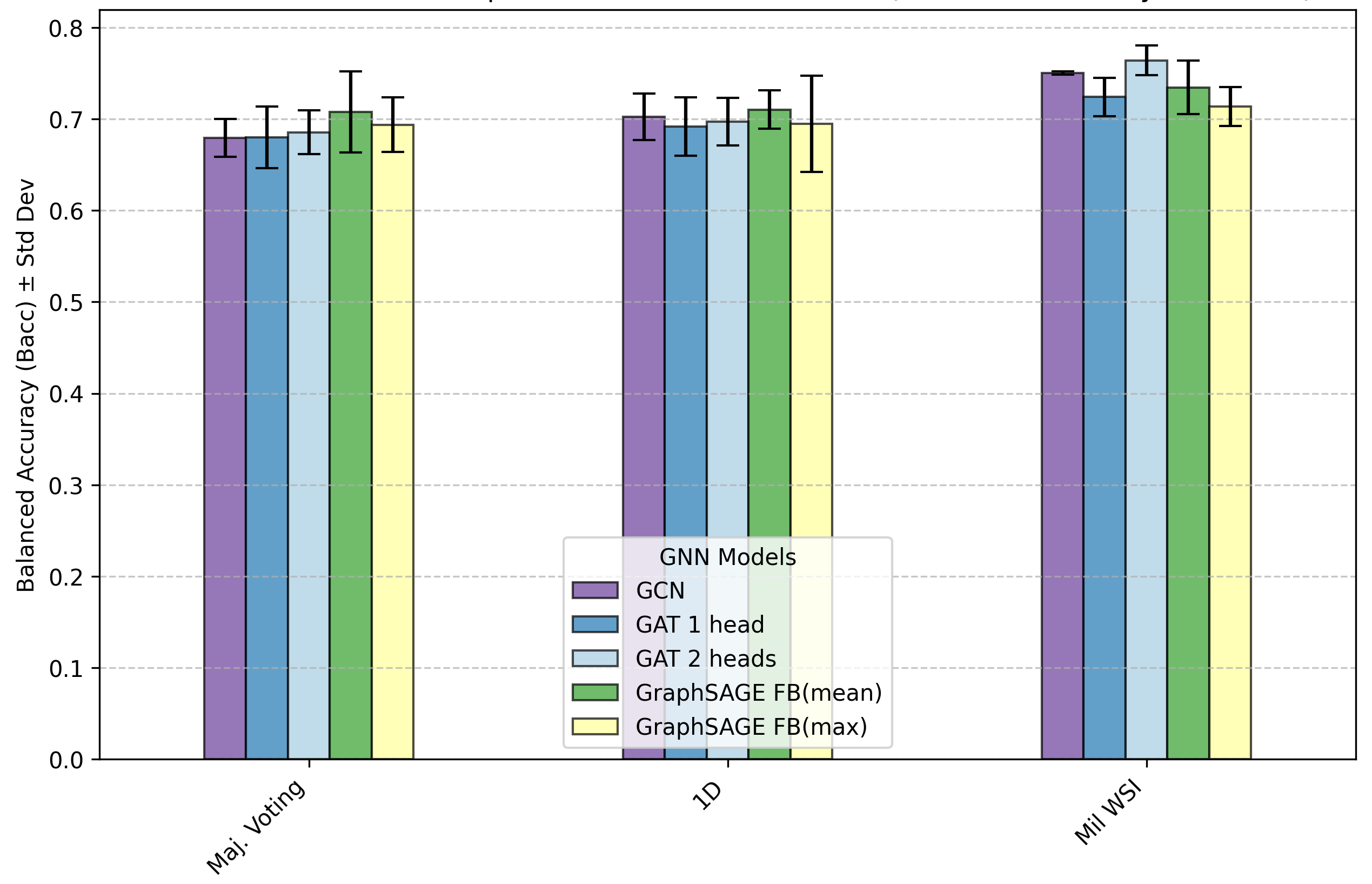}
        \caption{Balanced Accuracy}
        \label{fig:specificity}
    \end{subfigure}

    \caption{Performance comparison of different GNN architectures under multiple strategies for survival prediction on MMIST-ccRCC.}
    \label{fig:patient_prediction}
\end{figure}

Since patients often have multiple WSIs, we evaluate the impact of patient-level aggregation by comparing the following strategies: (i) One-Dominance (1D): Selecting the most confident WSI prediction for each patient; (ii) Majority Voting (MV): Aggregating WSI predictions by assigning the most frequently predicted class as the final patient label; and (iii) MIL-Based Slide Selection: Automatically selecting the most representative WSI per patient. Figure \ref{fig:patient_prediction} summarizes the results of these three strategies.

\subsubsection{MIL: Improved Generalization with Higher Variability}
MIL-based WSI selection improves GCN's balanced accuracy from $0.7023 \pm 0.0255$ to $0.7505 \pm 0.0018$, stabilizing predictions by filtering out less informative WSIs. GAT with 2 heads achieves the highest recall ($0.8533 \pm 0.0570$) and balanced accuracy ($0.7642 \pm 0.0162$), but at the cost of lower specificity ($0.6750 \pm 0.0468$), highlighting a trade-off between sensitivity and false positives. GAT with 1 head achieves a more stable performance with recall ($0.6857 \pm 0.0319$), specificity ($0.7625 \pm 0.0468$), and balanced accuracy ($0.7241 \pm 0.0211$), indicating that while multi-head attention emphasizes crucial features, it also introduces variability. These results are represented in Figure \ref{fig:patient_prediction}. 

Aditionally, compared to MMIST-ccRCC's MIL-based CNN performance in \cite{ccRCC} ($0.69$ balanced accuracy), our best GNN model (GAT 2 heads with MIL selection) achieves a notable improvement ($0.76$ balanced accuracy). This underscores the advantages of GNNs in capturing spatial dependencies and integrating histopathological features, reinforcing their superiority over CNN-based approaches.

\subsubsection{One-Dominance (1D): Stable but Over-Conservative}

The 1D strategy prioritizes specificity, mimicking a conservative clinical decision-making approach where only the most confident WSI prediction is considered. This leads to a lower standard deviation in specificity, $\pm 0.00637$ for GCN and $\pm 0.00729$ for GAT 2 heads, indicating greater stability but also a higher likelihood of discarding valuable tumor variability that could contribute to more comprehensive patient characterization. These results are also represented in Figure \ref{fig:patient_prediction}. 

\subsubsection{Majority Voting (MV): A Middle-Ground Trade-Off}
 MV balances recall and specificity but introduces moderate performance fluctuations. MV exhibits slightly lower recall stability than MIL selection, with higher recall variance ($\pm 0.0584$ for GAT 1 head). While it improves balanced accuracy, its inconsistent performance suggests it serves as a trade-off between the high sensitivity of MIL selection and the stability of 1D. These results are also represented in Figure \ref{fig:patient_prediction}. 
 

\subsubsection{Evaluating MIL Selection with DX Slides} We further assess MIL selection quality by replacing 385 TCGA slides with their corresponding DX samples while keeping CPTAC unchanged. Performance differences were minor, with balanced accuracy remaining stable ($0.7664 \pm 0.0162$ vs. $0.7642 \pm 0.0162$). Specificity improved ($0.7500 \pm 0.0559$ vs. $0.6750 \pm 0.0468$), while recall decreased ($0.7829 \pm 0.0641$ vs. $0.8533 \pm 0.0570$). This suggests that multiple WSIs may contain valuable prognostic information and that future work should explore integrating multiple WSIs instead of selecting a single representative slide.

\subsection{Conclusion}
Our study identifies GraphSAGE (mean pooling) and GAT (2-head attention) as the most effective GNN architectures for WSI-based survival prediction, achieving superior balanced accuracy and generalization. MIL-based selection significantly enhances patient-level predictions, particularly recall, effectively capturing aggressive tumor characteristics essential for survival analysis.

GNNs offer a robust alternative to CNNs, preserving spatial dependencies and modeling tumor heterogeneity. Among aggregation strategies, 1D provides stable but conservative predictions, MV balances recall and specificity with moderate variability, and MIL-based selection delivers the best generalization, although with minor fluctuations in specificity.

By surpassing CNN-based MIL benchmarks, GNNs set a new standard for survival prediction in renal cell carcinoma. Future work should explore multi-WSI integration to further optimize survival models by leveraging complementary tumor information across different slides.

\begin{credits}
\subsubsection{\ackname} 
We would like to express our sincere gratitude to Professor Carlos Santiago for his invaluable guidance and support throughout this work. 
The results shown in this work used datasets collected from MMIST: https://Multi-Modal-IST.github.io/.  This work was supported by LARSyS funding 
DOI: $(10.54499$/ $LA$/ $P$/ $0083$/ $2020$, $10.54499$/$UIDP$/$50009$/$2020$, and $10.54499$$/UIDB$$/50009$/$2020$), the FCT-funded scholarship $2023.02043.BDANA$ and project Center for Responsible AI [PRR - C645008882-00000055].

\subsubsection{\discintname}
We do not have any competing interest to declare.
\end{credits}
%
%
%

\begin{thebibliography}{8}

\bibitem{Zhang}
Zhang, J., et al. (2022, September). Gigapixel whole-slide images classification using locally supervised learning. In International conference on medical image computing and computer-assisted intervention (pp. 192-201). Cham: Springer Nature Switzerland.

\bibitem{tcga}
Clark, K., et al. (2013). The Cancer Imaging Archive (TCIA): maintaining and operating a public information repository. Journal of digital imaging, 26, 1045-1057.

\bibitem{ccRCC}
Mota, T., et al.  (2024). MMIST-ccRCC: A Real World Medical Dataset for the Development of Multi-Modal Systems. In Proceedings of the IEEE/CVF Conference on Computer Vision and Pattern Recognition (pp. 2395-2403).

\bibitem{Fatima}
Fatima, S., et al. (2023). A comprehensive review on multiple instance learning. Electronics, 12(20), 4323.

\bibitem{Tarkhan}
Tarkhan, A., et al. (2022, March). Attention-based deep multiple instance learning with adaptive instance sampling. In 2022 IEEE 19th International Symposium on Biomedical Imaging (ISBI) (pp. 1-5). IEEE.

\bibitem{Ilse}
Ilse, M., et al.  (2018, July). Attention-based deep multiple instance learning. In International conference on machine learning (pp. 2127-2136). PMLR.

\bibitem{Li}
Li, B., et al. (2021). Dual-stream multiple instance learning network for whole slide image classification with self-supervised contrastive learning. In Proceedings of the IEEE/CVF conference on computer vision and pattern recognition (pp. 14318-14328).

\bibitem{Lii}
Li, R., et al. (2018, September). Graph CNN for survival analysis on whole slide pathological images. In International Conference on Medical Image Computing and Computer-Assisted Intervention (pp. 174-182). Cham: Springer International Publishing.

\bibitem{Wang}
Wang, Z., Ma, J., Gao, Q., Bain, C., Imoto, S., Liò, P., ... \& Song, J. (2024). Dual-stream multi-dependency graph neural network enables precise cancer survival analysis. Medical Image Analysis, 97, 103252.

\bibitem{Zhao}
Zhao, Y., et al. (2020). Predicting lymph node metastasis using histopathological images based on multiple instance learning with deep graph convolution. In Proceedings of the IEEE/CVF conference on computer vision and pattern recognition (pp. 4837-4846).

\bibitem{Raju}
Raju, A., et al. (2020). Graph attention multi-instance learning for accurate colorectal cancer staging. In Medical Image Computing and Computer Assisted Intervention–MICCAI 2020: 23rd International Conference, Lima, Peru, October 4–8, 2020, Proceedings, Part V 23 (pp. 529-539). Springer International Publishing.

\bibitem{Shi}
Shi, J., et al. (2023). A structure-aware hierarchical graph-based multiple instance learning framework for pt staging in histopathological image. IEEE Transactions on Medical Imaging, 42(10), 3000-3011.

\bibitem{Zhou}
Zhou, Y., et al. (2019). Cgc-net: Cell graph convolutional network for grading of colorectal cancer histology images. In Proceedings of the IEEE/CVF international conference on computer vision workshops (pp. 0-0).

\bibitem{Anklin}
Anklin, V., et al. (2021). Learning whole-slide segmentation from inexact and incomplete labels using tissue graphs. In Medical Image Computing and Computer Assisted Intervention–MICCAI 2021: 24th International Conference, Strasbourg, France, September 27–October 1, 2021, Proceedings, Part II 24 (pp. 636-646). Springer International Publishing.

\bibitem{Chen}
Chen, R., et al.  \& Mahmood, F. (2021). Whole slide images are 2d point clouds: Context-aware survival prediction using patch-based graph convolutional networks. In Medical Image Computing and Computer Assisted Intervention–MICCAI 2021: 24th International Conference, Strasbourg, France, September 27–October 1, 2021, Proceedings, Part VIII 24 (pp. 339-349). Springer International Publishing.

\bibitem{Ritap}
Pereira, R., et al.  (2025). The Role of Graph-based MIL and Interventional Training in the Generalization of WSI Classifiers. arXiv preprint arXiv:2501.19048.

\bibitem{Liii}
Ruoyu Li, et al. Graph cnn for survival analysis on
whole slide pathological images. In Medical Image Computing and Computer
Assisted Intervention – MICCAI 2018, pages 174–
182, Cham, 2018. Springer International Publishing. ISBN 978-3-030-00934-2.

\bibitem{clam}
Lu, M., et al. (2021). Data-efficient and weakly supervised computational pathology on whole-slide images. Nature biomedical engineering, 5(6), 555-570.

\bibitem{GCN}
Kipf, T. et al. (2016). Semi-supervised classification with graph convolutional networks. arXiv preprint arXiv:1609.02907.

\bibitem{GAT}
Veličković, P., et al. (2017). Graph attention networks. arXiv preprint rXiv:1710.10903.

\bibitem{GraphSage}
Hamilton, W., et al. (2017). Inductive representation learning on large graphs. Advances in neural information processing systems, 30.



\end{thebibliography}
%

\end{document}